\documentclass[prc,amsmath,showpacs,twocolumn,floatfix,unsortedaddress]{revtex4}
\usepackage{graphics}

\usepackage{amssymb,amsfonts,amsmath,amsthm,amsbsy,mathrsfs,makeidx,textcomp,enumerate,graphicx}
\usepackage{dcolumn}
\usepackage[pdftex,dvipsnames,usenames]{color}		

\def\beq{\begin{equation}}
\def\eeq{\end{equation}}
\def\bea{\begin{eqnarray}}
\def\eea{\end{eqnarray}}
\newcommand*{\tabref}[1]{Table~\ref{tbl:#1}}
\newcommand*{\tablab}[1]{\label{tbl:#1}}
\renewcommand*{\eqref}[1]{Eq.~(\ref{eq:#1})}
\newcommand*{\eqlab}[1]{\label{eq:#1}}
\newcommand*{\figref}[1]{Fig.~\ref{fig:#1}}
\newcommand*{\figlab}[1]{\label{fig:#1}}

\newcommand*{\secref}[1]{Section~\ref{sec:#1}}
\newcommand*{\seclab}[1]{\label{sec:#1}}

\def\VYP#1#2#3{{\bf #1}, #3 (#2)}  
\def\NP#1#2#3{Nucl.~Phys.\ \VYP{#1}{#2}{#3}}
\def\NPA#1#2#3{Nucl.~Phys.\ \VYP{A#1}{#2}{#3}}

\def\PLB#1#2#3{Phys.~Lett.\ B~\VYP{#1}{#2}{#3}}
\def\PR#1#2#3{Phys.~Rev.~\VYP{#1}{#2}{#3}}
\def\PRL#1#2#3{Phys.~Rev.~Lett.\ \VYP{#1}{#2}{#3}}

\def\ZP#1#2#3{Z.~Phys.\ \VYP{#1}{#2}{#3}}

\def\etal{ {\it et al.}}
\def\NDS#1#2#3{Nuclear Data Sheets\ ~\VYP{#1}{#2}{#3}}
\def\PRC#1#2#3{Phys.~Rev.\ C~\VYP{#1}{#2}{#3}}
\def\JPG#1#2#3{J.~Phys.\ G~\VYP{#1}{#2}{#3}}
\def\JPC#1#2#3{J.~Phys.\ C~\VYP{#1}{#2}{#3}}
\def\Sr#1#2#3{Phys.~Scr.\ ~\VYP{#1}{#2}{#3}}
\def\PJPV#1#2#3{Pramana-J.~Phys.\ ~\VYP{#1}{#2}{#3}}
\def\JPL#1#2#3{J.~Phys.\ ~\VYP{#1}{#2}{#3}}

\newcommand{\Omit}[1]{}

\begin{document}

\title{A description of odd mass Xe and Te isotopes in the Interacting Boson-Fermion Model}
\author{S.~Abu-Musleh$^{1,2}$,  H.M.~Abu-Zeid$^2$ and O.~Scholten$^3$}
\affiliation{$^1$ 
 National Center Of Research  Palestine   -   Gaza\\
 $^2$ Phys. Dep., Faculty of Women for Art, Science \& Education, Ain Shams University, Cairo, Egypt\\
 $^3$ Kernfysisch Versneller Instituut, University of Groningen, 9747 AA, Groningen, The Netherlands}
\date{\today}

\begin{abstract}

Recent interest in spectroscopic factors for single-neutron transfer in low-spin states of the even-odd Xenon $^{125,127,129.131}$Xe and even-odd Tellurium, $^{123,125,127,129,131}$Te isotopes stimulated us to study these isotopes  within the frame work of the Interacting Boson-Fermion Model. The fermion that is coupled to the system of bosons is taken to be in the positive parity $3s_{1/2}$, $2d_{3/2}$, $2d_{5/2}$, $1g_{7/2}$ and in the negative $1h_{11/2}$ single-particle orbits, the complete 50-82 major shell. The calculated energies of low-spin energy levels of the odd isotopes are found to agree well with the experimental data. Also B(E2), B(M1) values and spectroscopic factors for single-neutron transfer are calculated and compared with experimental data.

\end{abstract}

 \maketitle

\section{Introduction}

The nuclei in the A=130 mass region show a transition from vibrational to tri-axial and for this reason have attracted much interest for nuclear structure calculations as they are a good test case for models.
It should also be noted that in this mass region there are several nuclei that decay by double beta decay. The possibility to use this decay mode for studying neutrino process increases the interest in their structure.

In recent papers~\cite{J99,Bu91,V00,H5,J84,H07}, detailed level schemes of $^{123,125,127,129,131}$Te are presented including new data on spectroscopic factors for the single-neutron stripping reactions as obtained from $(d,p)$ reactions. A notable calculation has appeared recently~\cite{Yos03} of nuclei in this mass region in the IBA-2 model for the purpose of predicting the double-beta decay matrix element.
In the present work we will focuss on the odd-neutron nuclei in this mass region and consider them in the framework of the Interacting Boson Model (IBM) and its extension to odd-mass isotopes, the Interacting Boson-Fermion Model~\cite{o01} (IBFM).

In a recent paper~\cite{Pascu10} a comprehensive and detailed calculation is presented for the Z=52-62 even-mass neutron-deficient nuclei in terms of the IBM framework. This calculation shows a change of structure from spectra close to the vibrational limit U(5) for Te isotopes evolving to almost tri-axial, near O(6), for the Xe isotopes.

Several theoretical and experimental~\cite{o02,m85,o04,o05,o06,o07,Y08} studies have been applied to level schemes and electromagnetic transition rates in different odd-mass Xe and Ba isotopes, in stark contrast with the rare attention given to spectroscopic factors which are very sensitive to details of wave functions and therefore can provide a fine test of the model.
In a recent work~\cite{Y08} the IBFM is used, in a in some sense very similar calculation as presented here, to obtain a description for positive parity levels in two odd-mass Xe-isotopes based on coupling the  $3s_{1/2}$, $2d_{3/2}$, $2d_{5/2}$, $1g_{7/2}$ single-particle levels to the even-mass cores. No results for negative parity levels, based on the $1h_{11/2}$ negative parity single-particle level, were presented.
In the present work we have set out to obtain a consistent description of positive and negative parity states of the even-odd  Tellurium isotopes, $^{123,125,127,129,131}$Te, and even-odd Xenon isotopes, $^{125,127,129,131}$Xe, which have 52,54 protons and 71-79 and 71-75 neutrons respectively. These nuclei are therefore described in the IBFM by the coupling of the degrees of freedom of a single neutron in the $3s_{1/2}$, $2d_{3/2}$, $2d_{5/2}$, $1g_{7/2}$, and  $1h_{11/2}$ single particle levels (the complete 50-82 major shell) to the even-even cores as described in~\cite{Pascu10}. This allows us, for the first time, to perform unified calculations for the positive and negative parity states of the even-odd Te and Xe isotopes using the same interaction strength for both parities.

The fact that the same coupling is used for positive and negative parity states allowed us to introduce another new aspect in the calculations. In the present calculations the quasi-particle energies and occupation probabilities are not obtained from a BCS calculation but adjusted to reach an optimal fit to the data. To do so realistically it is necessary to consider also the spectroscopic factors in the calculation as these depend very sensitively on the s.p.\ occupation probabilities.

In \secref{IBFM}, the details of the model will be described. In Sec. III and IV  the results on energy levels and electromagnetic transition rates will be presented, and in V the results of spectroscopic factors. This final step is parameter free and provides a unique test of the model. Our calculations in good agreement with the a valuable data.

\section{The Interacting Boson-Fermion Model (IBFM).}\seclab{IBFM}

In the IBFM~\cite{IP91}, odd-A nuclei are described by coupling of the effective degrees of freedom of the low-energy quasi-particle levels to a core described in the IBM. The total Hamiltonian is thus written as the sum of three parts,
\begin{equation}
H=H_{B} + H_{F} + V_{BF} \; , \eqlab{H-IBFA}
\end{equation}
where, $H_{B}$ is the usual IBM-1 Hamiltonian~\cite{Iach87}, describing the even-even core, $H_{F}$ is the fermion Hamiltonian  containing only one-body terms (since the effective degrees of freedom of only a single fermion is coupled to the bosons), and $V_{BF}$ is the boson-fermion interaction describing the interaction between the fermion and the bosons of the even-even core nucleus~\cite{Olaf85}.

The one-body term is written as
\begin{equation}
H_{F} =\sum_{jm}\varepsilon_{j}  a_{jm}^\dagger \; a_{jm} \; ,
\end{equation}
where $\varepsilon_{j}$ denotes the quasi-particle energies and $a_{jm}^\dagger$ and $a_{jm} $ are the creation and annihilation operators for the quasi-particle in the eigenstate $\left| jm \right\rangle$.

 
\begin{table*}[htb!]
\caption{The IBFM parameters as used in $^{125,127,129,131}$Xe calculations compared to published values.
All parameters are in MeV except for $\chi$ which is dimensionless.\tablab{Compare-Xe}}
 \begin{tabular}{|c||c|c|c||c|c|c||c|c|c||c|}  
  \hline
  & \multicolumn{3}{c||}{$\Lambda_0$} & \multicolumn{3}{c||}{$\Gamma_0$} & \multicolumn{3}{c||}{$A_0$} & -$\chi$ \\
  \hline & This & \multicolumn{2}{p{2.5cm}||}{\centering Published work}& This & \multicolumn{2}{p{2.5cm}||}{Published work}& This & \multicolumn{2}{p{2.5cm}||}{Published work}& This \\
  Parity  &+ \& -&   +   & -    &+ \& -& +    & -     &+ \& -& +     &  -   &+ \& -\\
  \hline
$^{125}Xe$& 1.41 & -0.54{\cite{o05}} & 0.07{\cite{o06}} & 0.25 & -0.04{\cite{o05}}& -0.038{\cite{o06}}& -0.15 & -     & -    & 1.24\\
          &      &  0.46{\cite{K12}} & 1.59{\cite{K12}} &      & 0.21{\cite{K12}} &  0.74{\cite{K12}} &       & -0.12{\cite{K12}} & -0.29{\cite{K12}}&\\
          &      & 0.41{\cite{Y08}}  &                  &      & 0.23{\cite{Y08}} &                   &       & -0.11{\cite{Y08}} &      &\\
  \hline
$^{127}Xe$& 1.4  &-0.44{\cite{o05}}  &-0.11{\cite{o06}} & 0.25 &-0.04{\cite{o05}} & -0.038{\cite{o06}}& -0.15 &   -   & -    & 1.33\\
          &      &  0.46{\cite{K12}} & 1.71{\cite{K12}} &      & 0.21{\cite{K12}} & 1.06{\cite{K12}}  &       &-0.12{\cite{K12}}  &-0.33{\cite{K12}} &\\
  \hline
$^{129}Xe$& 1.4  & -0.28{\cite{o05}} &-0.28{\cite{o06}} & 0.25 &-0.04{\cite{o05}} &-0.054{\cite{o06}} & -0.15 &  -    & -    &1.28\\
          &      & 0.4{\cite{K12}}   & 2.21{\cite{K12}} &      & 0.24{\cite{K12}} & 1.23{\cite{K12}}  &       & -0.17{\cite{K12}} & -0.3{\cite{K12}} &\\
          &      & 0.37{\cite{Y08}}  &                  &      & 0.25{\cite{Y08}} &                   &       & -0.16{\cite{Y08}} &      &\\
  \hline
$^{131}Xe$& 1.4  &-0.28{\cite{o05}}  &-0.29{\cite{o06}} & 0.25 &-0.04{\cite{o05}} &-0.083{\cite{o06}} &-0.15  & -     & -    &1.28\\
          &      & 0.29{\cite{K12}}  & 2.02{\cite{K12}} &      & 0.39{\cite{K12}} & 1.18{\cite{K12}}  &       & -0.19{\cite{K12}} &-0.24{\cite{K12}} &\\
          &      & 0.37{\cite{Y08}}  &                  &      & 0.25{\cite{Y08}} &                   &       & -0.16{\cite{Y08}} &      &\\
  \hline
  \end{tabular}
\end{table*}

\begin{table}[htb!]
\caption{Same as \tabref{Compare-Xe} but for the $^{123,125,127,129,131}$Te isotopes.\tablab{Compare-Te}}
 \begin{tabular}{|c||c|c|c|c||c|c|c|c||} 
  \hline
 & \multicolumn{4}{c||}{This work} & \multicolumn{4}{c||}{Published work}\\
  \hline
  & $\Lambda_0$& $\Gamma_0$& $-A_0$& -$\chi$ & ref& $\Lambda_0$& $\Gamma_0$& $-A_0$\\ \hline
$^{123}Te$ & 1.4 & 0.25 & 0.15 &1.239 & {\cite{V00}} & 0.95 & 0.2  & 0.15 \\
$^{125}Te$ & 1.4 & 0.25 & 0.15 &1.33  & {\cite{J99}} & 0.98 & 0.36 & 0.21\\
$^{127}Te$ & 1.4 & 0.25 & 0.15 &1.284 & {\cite{H5}}  & 0.95 & 0.2  & 0.24\\
$^{129}Te$ & 1.4 & 0.25 & 0.15 &1.284 & {\cite{H07}} & 0.95 & 0.2  & 0.20\\
$^{131}Te$ & 1.4 & 0.25 & 0.15 &1.284 & {\cite{J84}} & 0.95 & 0.35 & 0.20\\
\hline
  \end{tabular}
\end{table}

\begin{table}[htb!]
   \caption{ \tablab{BCS-param-Xe}
Occupation probabilities and quasi-particle energies for the $3s_{1/2}$, $2d_{3/2}$, $3d_{5/2}$, $1g_{7/2}$ and the $1h_{11/2}$ single particle orbits as used in the calculation of the $^{125,127,129,131}$Xe isotopes.}
\begin{tabular}
{|c||l|l||l|l||l|l||l|l||} \hline
 & \multicolumn{2}{c||}{$^{125}$Xe} &\multicolumn{2}{c||}{$^{127}$Xe} &\multicolumn{2}{c||}{$^{129}$Xe} &\multicolumn{2}{c||}{$^{131}$Xe}\\ \hline
$j$ & $v_j^2 $ & $\varepsilon_{j} $ & $v_j^2 $ & $\varepsilon_{j} $ & $v_j^2 $ & $\varepsilon_{j} $ & $v_j^2 $ & $\varepsilon_{j} $ \\
 \hline
$3s_{1/2}$ & .49 & 0.0 &     .49 & 0.0  &       .49 & 0.0     & .49 & 0.09 \\\hline
$2d_{3/2}$ & .52 & 0.27&     .52 & 0.30  &      .52 & 0.04    & .52 & 0.0  \\ \hline
$3d_{5/2}$ & .80 & 0.47&     .95 & 0.90       & .95 & 0.96    & .95 & 1.05 \\\hline
$1g_{7/2}$ & .93 & 0.55 &    .93 & 0.58      &  .93 & 0.85    & .93 & 0.94 \\\hline
$1h_{11/2}$& .65 & 0.24    & .58 & 0.20      &  .65 & 0.14   &  .65 & 0.07 \\\hline
\end{tabular}
\end{table}

\begin{table}[htb!]
   \caption{ \tablab{BCS-param-Te}
Same as \tabref{BCS-param-Xe} but for the $^{123,125,127,129,131}$Te isotopes.}
\begin{tabular}
{|c||l|l||l|l||l|l||l|l||l||l||} \hline
 & \multicolumn{2}{c||}{$^{123}$Te } &\multicolumn{2}{c||}{$^{125}$Te} &\multicolumn{2}{c||}{$^{127}$Te }  & \multicolumn{2}{c||}{$^{129}$Te } &\multicolumn{2}{c||}{$^{131}$Te }  \\ \hline
$j$ & $v_j^2 $ & $\varepsilon_{j} $ & $v_j^2 $ & $\varepsilon_{j} $ & $v_j^2 $ & $\varepsilon_{j} $  & $v_j^2 $ & $\varepsilon_{j} $ & $v_j^2 $ & $\varepsilon_{j} $\\
 \hline
$3s_{1/2}$ & .49 & 0.00 &     .49 & 0.0&      .49 &0.06  & .52 & 0.21 &   .52 & 0.30   \\\hline
$2d_{3/2}$ & .52 & 0.27&      .52 & 0.08&     .52 & 0.00 & .60 & 0.00&    .60 & 0.00   \\ \hline
$3d_{5/2}$ & .80 & 0.57&      .95 & 0.86    & .95 &0.92  & .98 & 0.61&    .98 & 0.68 \\ \hline
$1g_{7/2}$ & .93 & 0.65&      .93 & 0.48 &    .93 &0.81  & .93 & 0.81&    .94 & 1.03  \\ \hline
$1h_{11/2}$& .58 & 0.14     & .45 & 0.03&     .58 &0.03  & .55 & 0.05   & .66 & 0.16    \\ \hline
\end{tabular}
\end{table}
 

The boson-fermion interaction $V_{BF}$ is described in terms of three contributions; i) a monopole interaction which is characterized by the parameter $A_0$, ii) a quadruple interaction~\cite{IP91,BS70} characterized by $\Gamma_{0}$, and iii) the exchange of a quasi particle with one of the two fermions forming a boson~\cite{Iac81}, characterized by $\Lambda_{0}$, and is a consequence of the Pauli principle for the quadrupole interaction between protons and neutrons~\cite{Olaf85,IT81}. The full expression is written as
\begin{eqnarray}
V_{BV} &=& \sum_j A_{j}  [(d^\dagger \tilde{d})^{(0)} (a_j^\dagger \tilde{a}_j)^{(0)} ]
 \nonumber \\
 && + \sum_{jj'} \Gamma_{jj'} [Q^{(2)} (a_{j}^\dagger \tilde{a}_{j} )^{(2)} ]_{0}^{(0)}
 \nonumber \\
 && + \sum_{jj'j''} \Lambda_{jj'}^{j''}  :[(d^\dagger \tilde{a}_{j} )^{(j'')}
   (a_{j'}^\dagger \tilde{d}_{j} )^{(j'')} ]_{0}^{(0)}: \; ,\eqlab{V-IBFA}
\end{eqnarray}
where :\,: denotes normal ordering whereby contributions that arise from commuting the operators are omitted and $\tilde{a}_{jm} = (-1)^{j-m} a_{j-m}$ which implies
$$\sqrt{2j+1} (a_j^\dagger \tilde{a}_j)^{(0)}= \sum_m (-1)^{j-m} a_{jm}^\dagger \tilde{a}_{j-m} 
=- \hat{n}_j \;. $$
The monopole interaction  plays a minor role in the actual calculations.

The dominant terms in \eqref{V-IBFA} are the second and the third terms, which arise from the quadrupole interaction.  The remaining parameters in \eqref{V-IBFA} can be related to the BCS occupation probabilities $v_{j}$ of the single-particle orbits,
\begin{eqnarray}
\Gamma_{jj'} &=& \sqrt{5} \Gamma_{0} (u_{j} u_{j'} -v_{j} v_{j'})Q_{jj'} \;, \\
 \Lambda_{jj'}^{j''} &=& -\sqrt{5} \Lambda_{0}
  [(u_{j'} v_{j''} +v_{j'} u_{j''} )Q_{j'j''} \beta_{j''j}
   \nonumber \\
 && +(u_{j'} v_{j''} +v_{j'} u_{j''} )Q_{j'j} \beta_{j'j''} ]/\sqrt{2j''+1} \; ,
\end{eqnarray}
where $Q_{j'j''}$ are single particle matrix elements of the quadruple operator and
\begin{equation}
\beta_{jj'} ={(u_{j} v_{j'} + v_{j} u_{j'} )}Q_{jj'} \;,
\end{equation}
are the structure coefficients of the $d$-boson deduced from microscopic considerations~\cite{Olaf85}. The occupation probabilities and the quasi-particle energy of the single-particle orbitals can in principle be obtained by solving the gap equations. In the present calculations we have taken them as free parameters in addition to the strengths $\Lambda_{0}$, $\Gamma_{0}$ and $A_{0}$ to obtain the best fit to the excitation energies.

\section{Excitation Energies}\seclab{Ex}

\begin{figure}[htb!]  \begin{center}
\fbox{\includegraphics[angle=0,width=0.45\textwidth]{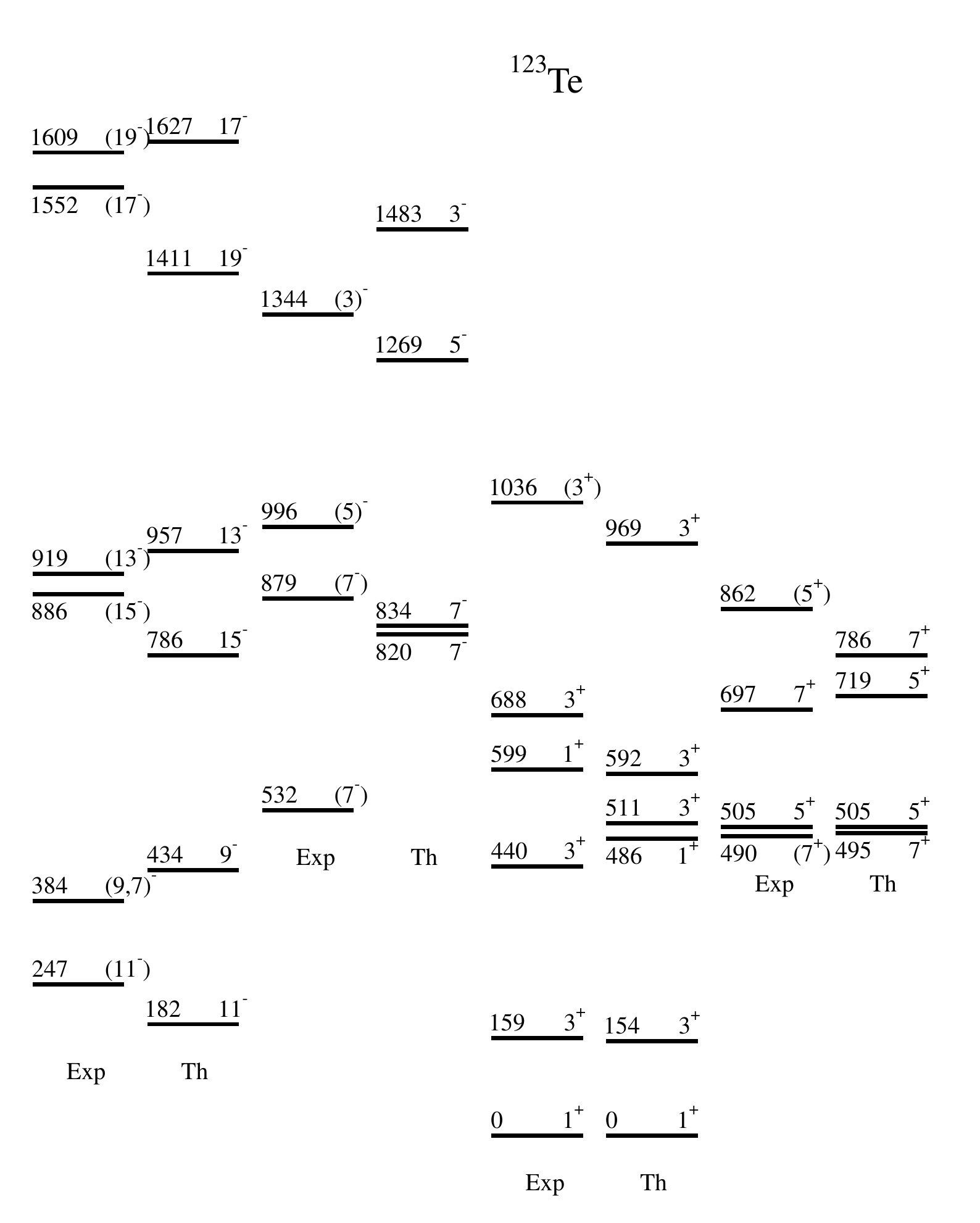}  }
\caption{Calculated energies for $^{123}$Te are compared to data~\cite{NNDC}. For each level the excitation energy in keV is given as well as the spin ($\times$ 2) and parity. \figlab{123Te-E}}
\end{center} \end{figure}

\begin{figure}[htb!]  \begin{center}
\fbox{\includegraphics[angle=0,width=0.45\textwidth]{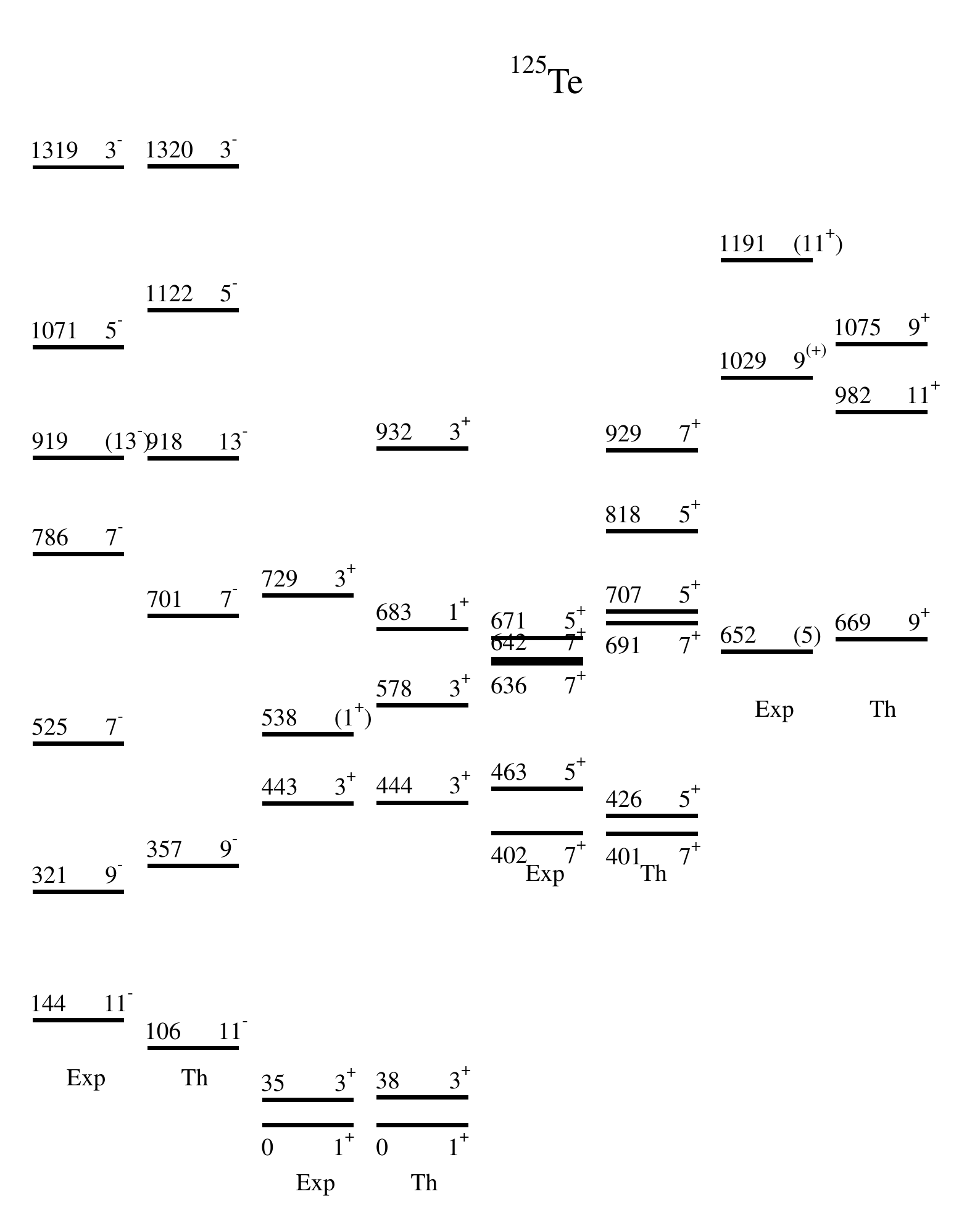} }
\caption{Same as in \figref{123Te-E} but for $^{125}$Te. \figlab{125Te-E}}
\end{center} \end{figure}

\begin{figure}[htb!]  \begin{center}
\fbox{\includegraphics[angle=0,width=0.45\textwidth]{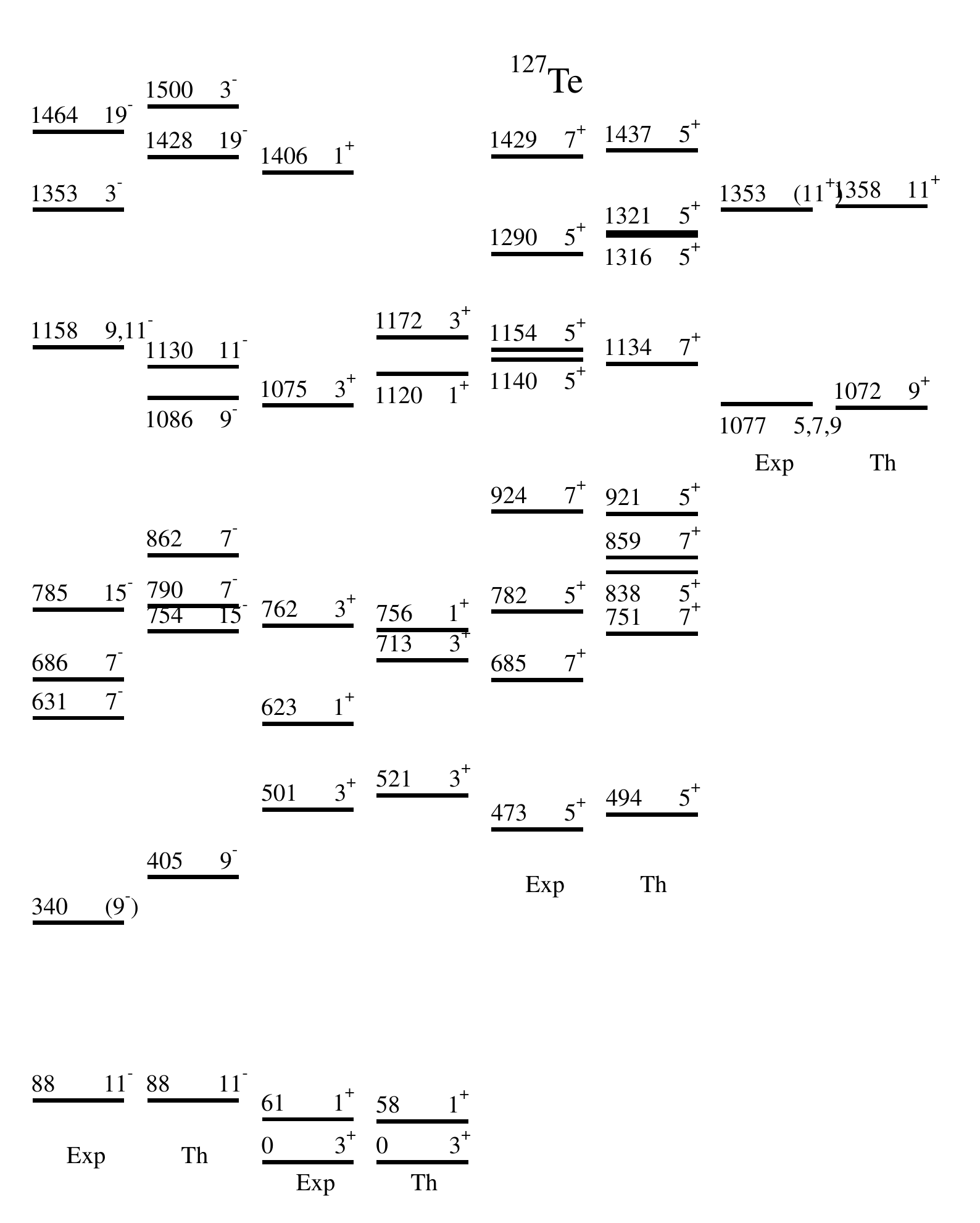} }
\caption{Same as in \figref{123Te-E} but for $^{127}$Te. \figlab{127Te-E}}.
\end{center} \end{figure}

\begin{figure}[htb!]  \begin{center}
\fbox{\includegraphics[angle=0,width=0.45\textwidth]{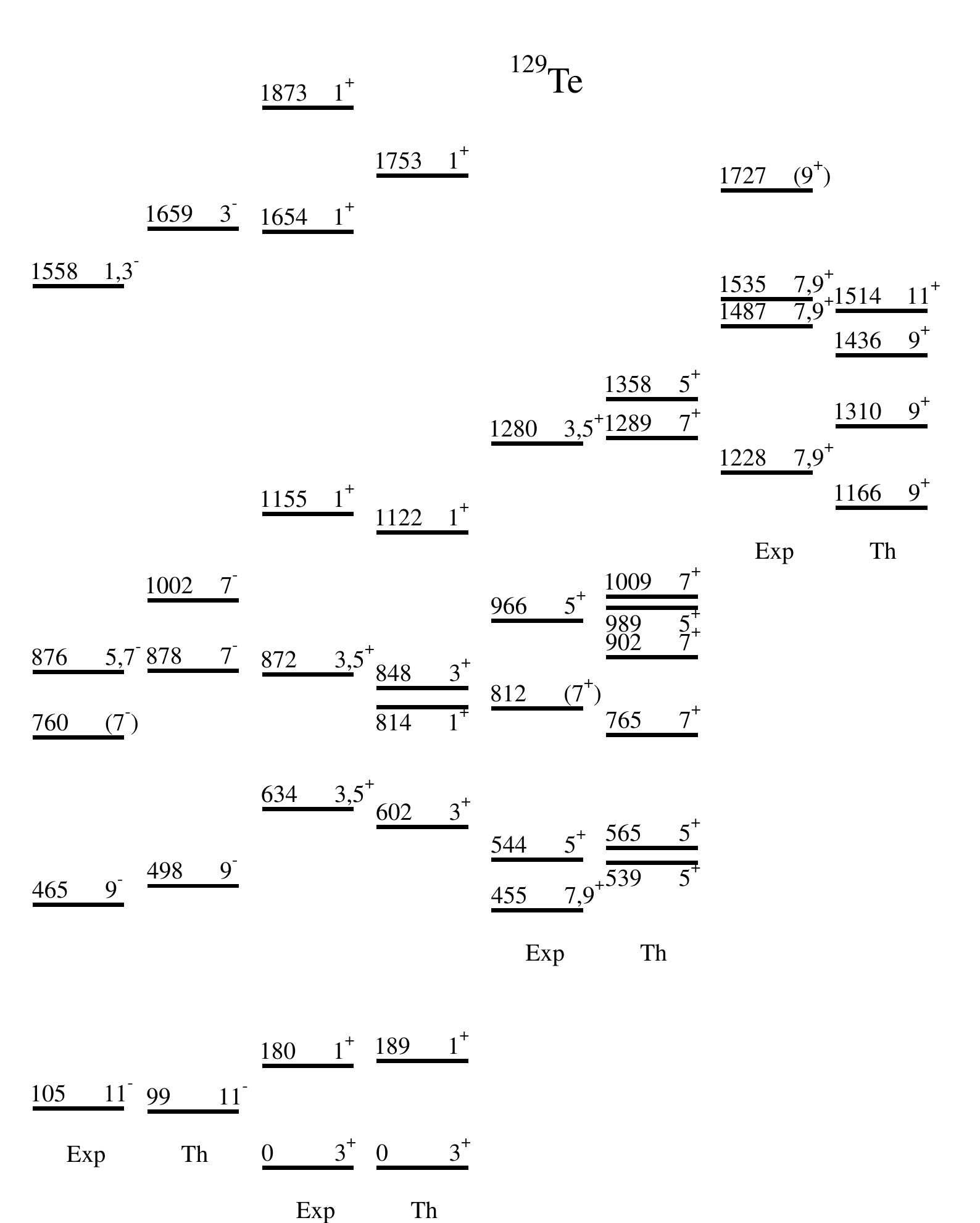} }
\caption{Same as in \figref{123Te-E} but for $^{129}$Te. \figlab{129Te-E}}.
\end{center} \end{figure}

\begin{figure}[htb!]  \begin{center}
\fbox{\includegraphics[angle=0,width=0.45\textwidth]{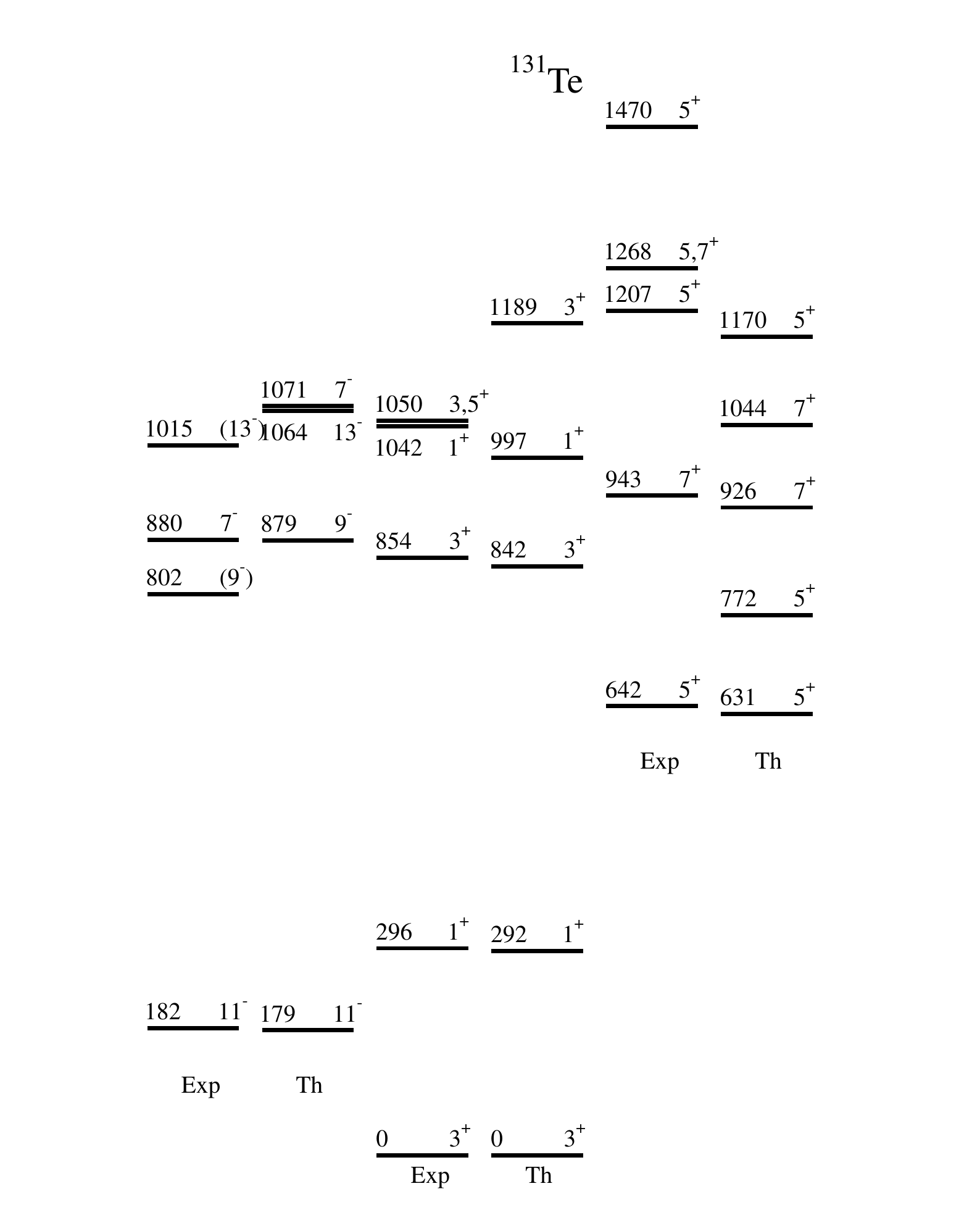} }
\caption{Same as in \figref{123Te-E} but for $^{131}$Te. \figlab{131Te-E}}.
\end{center} \end{figure}

The Hamiltonian, \eqref{H-IBFA} was diagonalized using the computer program ODDA~\cite{odda80} in which the IBFM parameters are identified as $A_0=BFM$, $\Gamma _0=BFQ$ and $\Lambda_0=BFE$. In the present study of the $^{125,127,129,131}$Xe and $^{123,125,127,129,131}$Te isotopes we have used the complete 50-82 major shell,  the $3s_{1/2}$, $2d_{3/2}$, $2d_{5/2}$, $1g_{7/2}$, and  $1h_{11/2}$ single particle orbits, for the odd-neutron quasi particle.
The quasi particle for these calculations is the fermion degree of freedom, describing a neutron hole, that is coupled to the bosons of the even-even may occupy. For the description of the even-even cores we have used the parameters as given in~\cite{Pascu10}. The use of the complete model space allows us, for the first time, to perform a comprehensive and unified calculations for the positive and negative parity states of the neutron-poor even-odd Te and Xe isotopes.

As part of our strategy to have a unified description we have tried, and succeeded, to keep the same values for all isotopes for the interaction strength of the Quadrupole, Exchange and Monopole forces, see \eqref{H-IBFA}.
We have been able to obtain good results for all isotopes under consideration using $\Lambda_{0}=1.4$\,MeV, $\Gamma_{0}=0.25$\,MeV and $A_{0}=-0.15$\,MeV.
The strength of the interaction is compared in \tabref{Compare-Xe} and \tabref{Compare-Te} with the values obtained in the literature. From these tables it will be clear that the values used in different works can differ greatly which stresses the need for the comprehensive approach we have taken. The differences are especially large for the strength of the exchange force although also in the strength of the quadrupole force substantial differences are seen. It should be noted that the effect of the strength of the quadrupole and exchange forces is modulated by the occupancy factors. By performing an overall fit to a larger series of isotopes and by including positive as well as negative parity states the freedom in the choice of the interaction strength is strongly limited.
The strength of the Monopole force does not have a very large effect on the results.

The quasi-particle energies and occupation probabilities were allowed to vary across the isotopes to get an optimal fit to excitation energies. At the same time we have kept an eye on single-particle transfer amplitudes (discussed in \secref{SPT}) as these are very sensitive to the occupation probabilities.
The used single-particle parameters are given in \tabref{BCS-param-Xe} for Xe and in \tabref{BCS-param-Te} for Te. Since only on the relative quasi-particle energies enter in the calculation of excitation energies we have set the lowest energy to zero. For Xe one observes that the energies as well as the occupation probabilities vary gradually over the mass range. For the Te-isotopes ones observes a similar gradual trend, with a minimum in the quasi=particle energies for the $g_{7/2}$ and $h_{11/2}$ orbits near $^{125}Te$ and a maximum for the $d_{5/2}$ close to $^{127}$Te. 
Based on the single particle energies as given in ~\cite{o06}, we have decided to include the $1h_{11/2}$ level and exclude the $1h_{9/2}$ orbit. With its much larger quasi-particle energy it is expected that the influence of the $1h_{9/2}$ orbit on the energy spectrum will be small.

\begin{figure}[htb!]  \begin{center}
\fbox{\includegraphics[angle=0,width=0.45\textwidth]{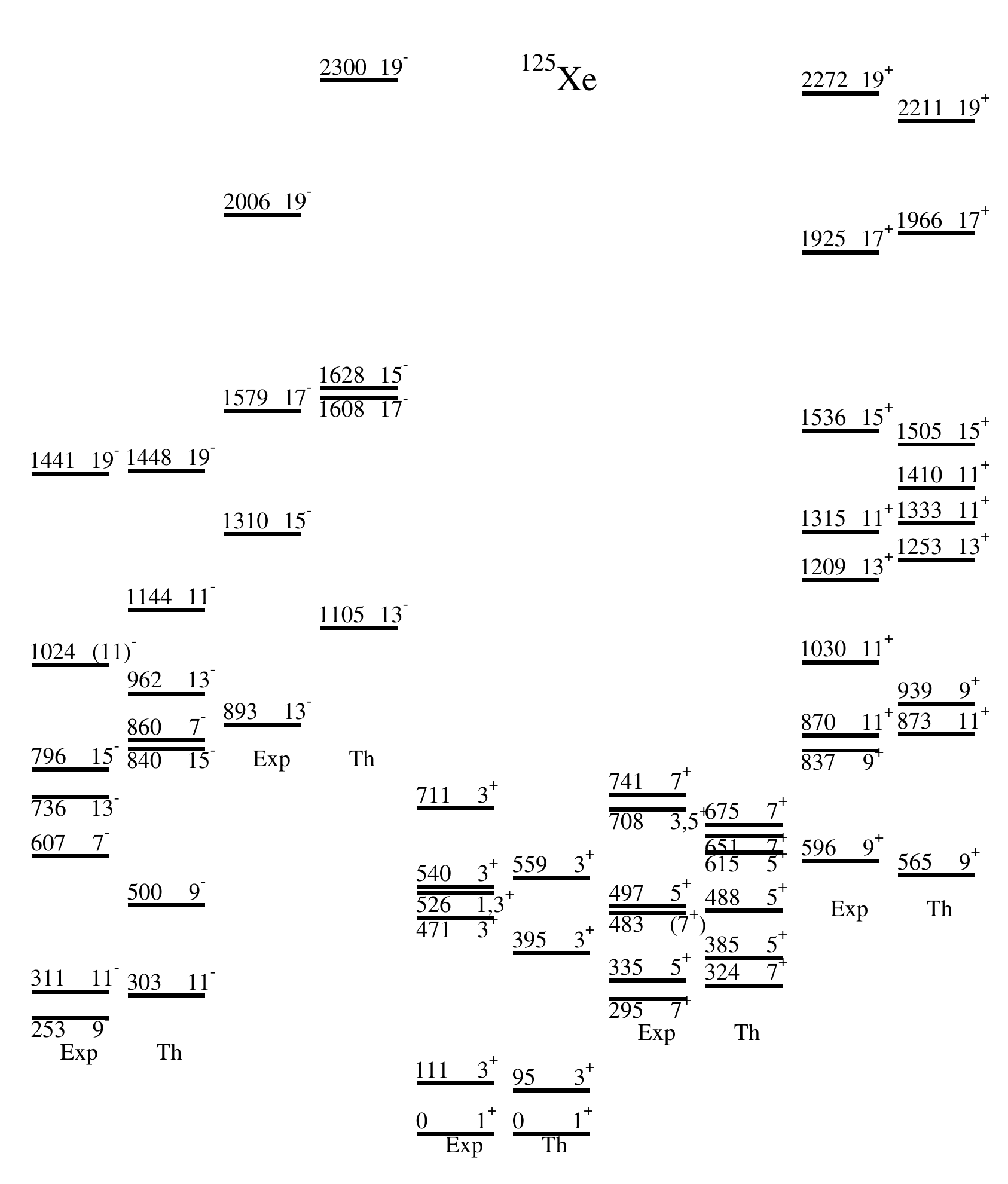}  }
\caption{Same as in \figref{123Te-E} but for $^{125}$Xe. \figlab{125Xe-E}}.
\end{center} \end{figure}

\begin{figure}[htb!]  \begin{center}
\fbox{\includegraphics[angle=0,width=0.45\textwidth]{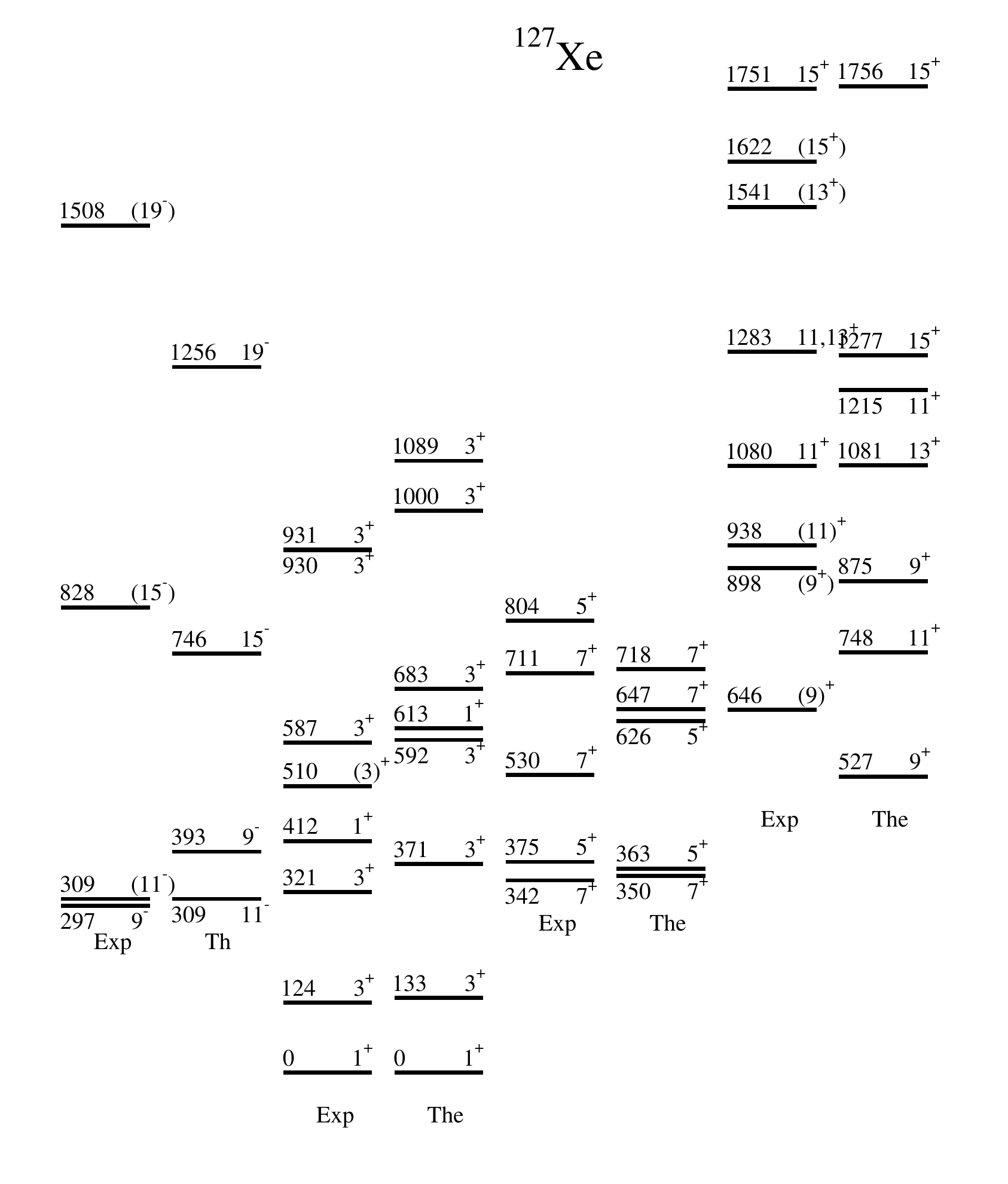} }
\caption{Same as in \figref{123Te-E} but for $^{127}$Xe. \figlab{127Xe-E}}.
\end{center} \end{figure}

\begin{figure}[htb!]  \begin{center}
\fbox{\includegraphics[angle=0,width=0.45\textwidth]{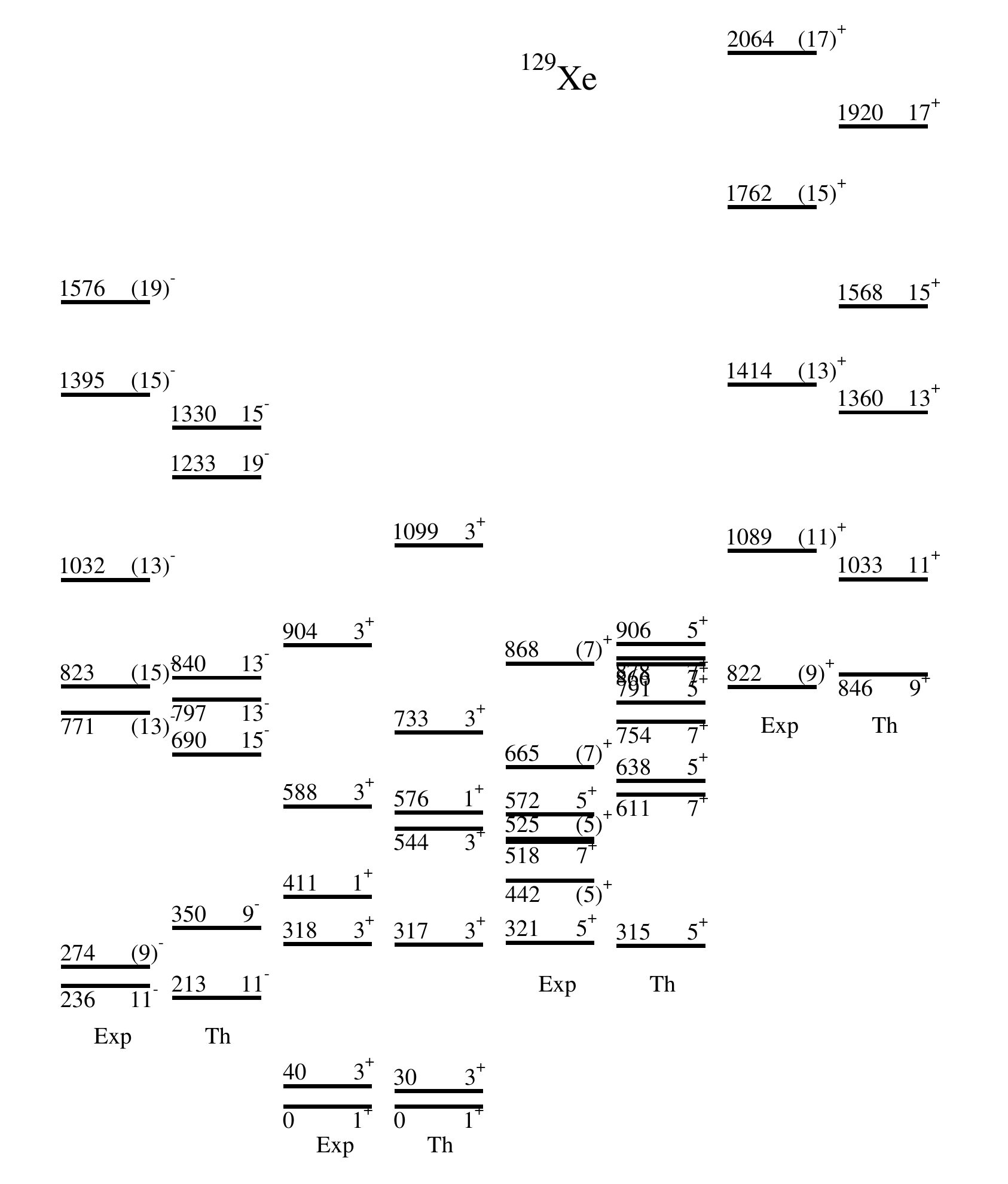} }
\caption{Same as in \figref{123Te-E} but for $^{129}$Xe. \figlab{129Xe-E}}.
\end{center} \end{figure}

\begin{figure}[htb!]  \begin{center}
\fbox{\includegraphics[angle=0,width=0.45\textwidth]{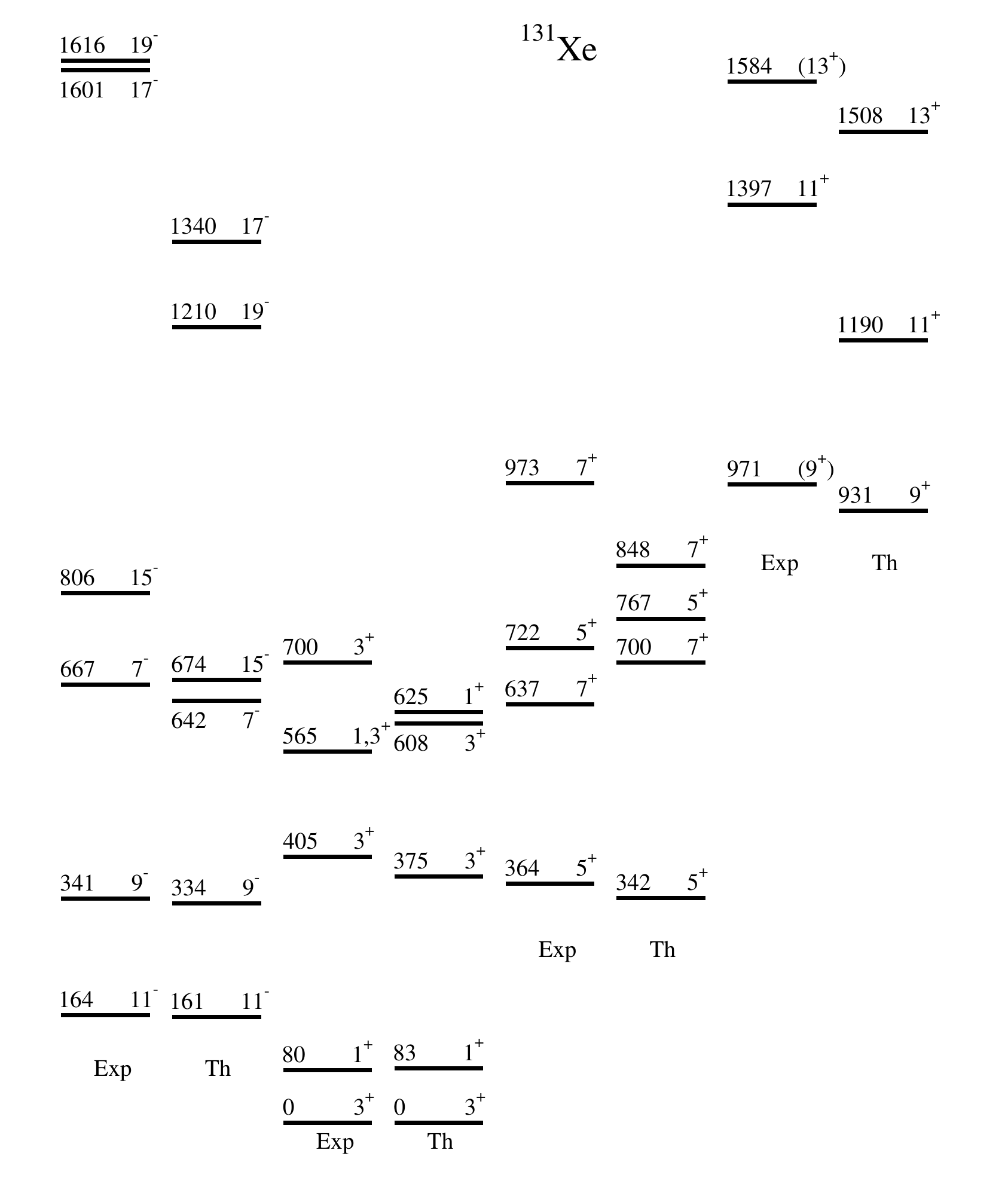} }
\caption{Same as in \figref{123Te-E} but for $^{131}$Xe. \figlab{131Xe-E}}.
\end{center} \end{figure}

The calculated excitation energies are compared with experimental data~\cite{NNDC} for the odd mass $^{125-131}$Xe and $^{123-131}$Xe isotopes in \figref{123Te-E}--\ref{fig:131Xe-E} for positive and negative parity states. In general a very good agreement is obtained for the spectra of all isotopes up to rather large excitation energies.
Since for these isotopes the levels can not really be arranged in clear bands that are connected by strong B(E2) transitions we have arranged the positive parity levels, where the level density is highest, by spin. This induces least bias in plotting and comparing the experimental levels with the calculations. We have tried to be complete in plotting the lowest few levels for each spin, unfortunately the spin assignments in the data are sometimes ambiguous.
In $^{125}$Xe, for example all levels have an ambiguous parity assignment. In the figures we have therefore indicated only the preferred parity assignments.

In general the level density for a given spin and parity is rather high which makes it difficult to make a clear link between the data and the calculated levels. However there are many quantitative features that that are well reproduced in the calculations. For example, in $^{127}$Te the high density of calculated $5/2^+$ levels agrees well with the data. For other isotopes, unfortunately, many of the spin assignments are not clear.

\section{Electromagnetic transition probabilities}\seclab{EM}

Electromagnetic transitions give a good test of the model wave functions where in particular the extent to which two wave functions have similar single particle components. In general, the electromagnetic transition operators are written as the sum of two terms, the first of which acts only on the boson part of the wave function and second only on the fermion part,
\begin{equation}
T^{(E2)} =e_b Q_{B}^{(2)}
 + e_{\nu} \sum_{jj''} Q_{jj'} (a_{j} \tilde{a}_{j})^{(2)} \;,
 \eqlab{T-E2-IBFA}
 \end{equation}
where $Q_B$ has been defined in \eqref{V-IBFA}, $Q_{j'j''}$ are single particle matrix elements of the quadruple operator, and $e_b$ and $e_{\nu}$ are the boson and fermion effective charges respectively.
In the calculations the boson effective charge $e_b$ was chosen such that it reproduces the experimental values for the even mass Xe and Te isotopes reasonably well with one value taken constant over the entire isotopic chain. This resulted in $e_b$=0.15\,eb for both Te and Xe. This value is in good agreement with the calculations presented in ref.~\cite{Pascu10} for the even-even cores. The fermion effective charge for Te, Xe is taken as $e_\nu$=0.15\,eb. It should be noted that the fermion effective charge has only a minor influence on the collective E2 transition strengths.

The M1 transition operator is given by
\bea
T^{(M1)} &=& {\sqrt{30/4\pi}}{g_{b}}\, ({{d^\dag} \times{ \tilde{d}})^{(1)}}
 \nonumber \\
 && - \sum_{jj'} g_{jj'} \sqrt{(j(j+1)(2j+1) \over 4\pi }\,( a_{j} \times\tilde{a}_{j})^{(1)} \;,
 \eqlab{T-M1-IBFA}
\eea
where,  $g_b=0.3$\,$\mu_N$ is the boson g-factor determined by the magnetic moment of levels in the even-even core and, $g_{jj'}$ is the single particle contribution which depend on $g_l$ and $g_s$ of the odd nucleon. In the actual calculations the computer program PBEM~\cite{Ph80} has been used.

 \begin{table}[htb!]
\begin{center}
  \caption{ \tablab{123Te-BE2-Wu}
Calculated and experimental values for B(E2) (right, in units of Wu) and B(M1) (left, in units of mWu) for $^{123}Te$. }
 \begin{tabular}{||c|c|c||c||c|c|c||}
  \hline
  \multicolumn{1}{|c|}{~\cite{V00}} &\multicolumn{1}{c|}{~\cite{Ohya93}}&\multicolumn{1}{c|}{This}& $^{123}$Te& \multicolumn{1}{c|}{This}   & \multicolumn{1}{|c|}{~\cite{Ohya93}}&
   \multicolumn{1}{|c|}{~\cite{V00}}\\
    \hline
 1.7&  23.$\pm$1.  & 0.51&  ${3\over 2}_{1}^+ \to {1\over 2}_{1}^+ $  & 2.5   & 2.5$\pm$.5 & 0.81\\
  \hline
 &&&                            ${7\over 2}_{1}^+ \to {3\over 2}_{1}^+ $  &4.6   &    & 9.7\\
  \hline
  &&&                           ${5\over 2}_{1}^+ \to {1\over 2}_{1}^+ $  & 0.11  &    &26. \\
  \hline
  10.&  11.$\pm$2.  & 75.9 &    ${5\over 2}_{1}^+ \to {3\over 2}_{1}^+ $  & 6.7   &  0.3$\pm$.5 &0.03 \\
  \hline
 7.6&  6.7$\pm$1.0  & 9.44&   ${3\over 2}_{2}^+ \to {3\over 2}_{1}^+ $  & 21.   & 2.2$\pm$1.6 & 0.02\\
  \hline
  0.2&  1.8$\pm$0.3  & 4.99&${3\over 2}_{2}^+ \to {1\over 2}_{1}^+ $  & 14.  & 29$\pm$4  &28.\\
  \hline
  \end{tabular}
\end{center}
\end{table}

\begin{table}[htb!]
\begin{center}
  \caption{ \tablab{125Te-BE2-Wu}
Same as \tabref{123Te-BE2-Wu} for $^{125}$Te. }
 \begin{tabular}{||c|c|c||c||c|c|c||}
  \hline
  \multicolumn{1}{|c|}{~\cite{J99}} &\multicolumn{1}{c|}{~\cite{CC97}}&\multicolumn{1}{c|}{This}& $^{125}$Te& \multicolumn{1}{c|}{This}   & \multicolumn{1}{c|}{~\cite{CC97}}&
   \multicolumn{1}{c|}{~\cite{J99}}\\
\hline
  0.9&  22.$\pm$1. & 0.217&      ${3\over 2}_{1}^+ \to {1\over 2}_{1}^+ $ & 0.22   & 10.2$\pm$2.2  & 1.9\\
  \hline
 1.1 &  1.3$\pm$0.1 & 1.57&      ${3\over 2}_{2}^+ \to {1\over 2}_{1}^+ $ & 22.  & 24.1$\pm$.9  & 9.5\\
  \hline
  &&&                           ${5\over 2}_{1}^+ \to {1\over 2}_{1}^+ $ & 8.1   & 13.9$\pm$0.6  &4.\\
  \hline
  &&&                           ${5\over 2}_{2}^+ \to {1\over 2}_{1}^+ $ & 0.03   & 11.8$\pm$0.6  &7.3 \\
 \hline
 11. &  51.$\pm$3. & 37.55&       ${5\over 2}_{2}^+ \to {3\over 2}_{1}^+ $ & 1.6 & 9.7$\pm$0.5 &4.9 \\
 \hline
 0.2 &  12.0$\pm$0.6 & 2.17&     ${5\over 2}_{1}^+ \to {3\over 2}_{1}^+ $ & 22.& 13.1$\pm$.8 &6.4 \\
  \hline
 1.0 &  1.9$\pm$.2 & 2.75&       ${3\over 2}_{2}^+ \to {3\over 2}_{1}^+ $ & 9.7 & 18.1$\pm$.8 & 1.\\
  \hline
  &&&                           ${7\over 2}_{1}^+ \to {3\over 2}_{2}^+ $ & 0.76  &   &0.3 \\
  \hline
  &&&                           ${7\over 2}_{1}^+ \to {3\over 2}_{1}^+ $ & 4.6  & 4.8$\pm$2.4 & 10.6\\
  \hline
  &&&                           ${7\over 2}_{2}^+ \to {3\over 2}_{1}^+ $  & 26.  & $>$2.3 & 0.8 \\
  \hline
  &&&                           ${7\over 2}_{2}^+ \to {3\over 2}_{2}^+ $  & 0.32 & $>$1.6 & 0.11 \\
  \hline
 0.0 &  1.1$\pm$0.6 & 1.05 &     ${7\over 2}_{1}^+ \to {5\over 2}_{1}^+ $  & 4.9  & 0.11 &1.7  \\
  \hline
 63. &  35.$\pm$2. & 22.9&       ${5\over 2}_{2}^+ \to {5\over 2}_{1}^+ $  & 0.13  & 6.2$\pm$1.7 &0.02 \\
  \hline
 110. &  7.1$\pm$0.4 & 4.97&     ${5\over 2}_{2}^+ \to {3\over 2}_{2}^+ $  & 2.9  & 95$\pm$6 & 0.25\\
  \hline
 2.5 &  3.8$\pm$0.1 & 0.34 &   ${9\over 2}_{1}^- \to {11\over 2}_{1}^- $  & 33.  &  30.2$\pm$1.6 &13.3 \\\hline
  &&&                           ${7\over 2}_{1}^- \to {11\over 2}_{1}^- $  & 29.  &  $>$9.4 &14. \\\hline
  &  &&                         ${7\over 2}_{1}^- \to {9\over 2}_{1}^- $  & 11.  &    &0.45 \\\hline
  &&&                           ${5\over 2}_{1}^- \to {9\over 2}_{1}^- $  & 14.  &     &5.7 \\
\hline
\end{tabular}
\end{center}
\end{table}

\begin{table}[htb!]
\begin{center}
  \caption{ \tablab{125Xe-BE2-wu}
 Calculated and experimental B(E2) values for $^{125}$Xe in units of Weisskopf units. }
 \begin{tabular}{||c||c|c|c||}
  \hline
   $^{125}$Xe& \multicolumn{1}{c|}{This }   & \multicolumn{1}{c|}{~\cite{Gamma}}&
   \multicolumn{1}{c|}{~\cite{Y08}} \\
    \hline
  ${3\over 2}_{1}^+ \to {1\over 2}_{1}^+ $  & 14.  & 100 $\pm$ 80  & 130.       \\
  \hline
  ${7\over 2}_{1}^+ \to {3\over 2}_{1}^+ $  & 3.5   & 0.32 $\pm$ 0.09 & 0.59 \\
  \hline
  ${5\over 2}_{1}^+ \to {1\over 2}_{1}^+ $  & 0.54  &  &  \\
  \hline
  ${5\over 2}_{1}^+ \to {3\over 2}_{1}^+ $  &8.0   & &  \\
  \hline
  ${11\over 2}_{1}^+ \to {7\over 2}_{1}^+ $  &19.   & 29 $\pm$ 7 & \\
  \hline
  ${13\over 2}_{1}^+ \to {9\over 2}_{1}^+ $  & 52.  & 57 $\pm$ 23 & \\
  \hline
  ${9\over 2}_{1}^+ \to {7\over 2}_{1}^+ $  &  86.   & 65 $\pm$ 19 &\\
  \hline
  ${15\over 2}_{1}^+ \to {11\over 2}_{1}^+ $  &  72.     & 43 $\pm$ 17 &\\
  \hline
  ${15\over 2}_{1}^- \to {11\over 2}_{1}^- $  &   56.   & 63 $\pm$ 14 &\\
  \hline
  ${19\over 2}_{1}^- \to {15\over 2}_{1}^- $  &   89.   & $>$ 46 &\\
  \hline
  \end{tabular}
\end{center}
\end{table}

\begin{table}[htb!]
\begin{center}
  \caption{ \tablab{127Xe-BE2-Wu}
 Calculated and experimental B(E2) values for $^{127}Xe$ in units of Wu.}
 \begin{tabular}{|c||c|c||}
  \hline
  $^{127}Xe$& \multicolumn{1}{c|}{This}   & \multicolumn{1}{c|}{~\cite{Gamma}}\\
   \hline
   ${3\over 2}_{1}^+ \to {1\over 2}_{1}^+ $   & 14.62   & 17$\pm$6  \\
  \hline
  ${7\over 2}_{1}^+ \to {3\over 2}_{1}^+ $   & 1.27   & 0.51$\pm$.09 \\
  \hline
  \end{tabular}
\end{center}
\end{table}

\begin{table}[htb!]
\begin{center}
  \caption{ \tablab{129Xe-BE2-Wu}
Calculated and experimental B(E2) values for $^{129}Xe$  in units of Wu. }
 \begin{tabular}{|c||c|c|c|c|c|c||}
  \hline
   $^{129}$Xe& \multicolumn{1}{c|}{This}   &\multicolumn{1}{c|}{~\cite{D78,A79,G74}}&\multicolumn{1}{c|}{~\cite{XW96}} &\multicolumn{1}{c|}{~\cite{J85}}&\multicolumn{1}{c|}{~\cite{m85}}& \multicolumn{1}{c|}{~\cite{o06}}  \\
     \hline
  ${3\over 2}_{1}^+ \to {1\over 2}_{1}^+ $  & 0.86&  8$\pm$4       &9.7 &1.9 && \\
 \hline
  ${3\over 2}_{2}^+ \to {3\over 2}_{1}^+ $  & 19.& 21.7$\pm$.9       &5.0 &3.5&7.3 &10. \\
 \hline
  ${3\over 2}_{2}^+ \to {1\over 2}_{1}^+ $  &23.&  5$\pm$15      &23. &32.  &27.3&18.9 \\
 \hline
  ${5\over 2}_{1}^+ \to {3\over 2}_{1}^+ $  & 32.& 56$\pm$7       &3. &27. &22. &22. \\
 \hline
  ${5\over 2}_{1}^+ \to {1\over 2}_{1}^+ $  & 10.&  20.8$\pm$2.1       &19. &10.5  &14.  &30.  \\
 \hline
  ${1\over 2}_{2}^+ \to {3\over 2}_{2}^+ $   & 3.9& 6.8$\pm$2.3       &10. & & & \\
  \hline
  ${5\over 2}_{2}^+ \to {1\over 2}_{1}^+ $  & 9.6 &           &1.1 &19. &18. &8.3\\
  \hline
  ${3\over 2}_{3}^+ \to {1\over 2}_{1}^+ $  & 14. &           &15. &1.5 &3. & 3.\\
  \hline
  ${3\over 2}_{4}^+ \to {1\over 2}_{1}^+ $  & 0.05 &           &3.5 &0.1 &0.3 & 1.4\\
  \hline
\end{tabular}
\end{center}
\end{table}

\begin{table}[htb!]
\begin{center}
  \caption{ \tablab{131Xe-BE2-Wu}
Calculated and experimental B(E2) values for  $^{131}Xe$ in units of Wu. }
 \begin{tabular}{|c||c|c|c|c|c|c||}
  \hline
  $^{131}$Xe& \multicolumn{1}{c|}{This}   &
   \multicolumn{1}{c|}{~\cite{D78,A79}}&\multicolumn{1}{c|}{~\cite{XW96}}& \multicolumn{1}{c|}{~\cite{J85}}& \multicolumn{1}{c|}{~\cite{m85}}& \multicolumn{1}{c|}{~\cite{o06}}  \\
     \hline
 ${1\over 2}_{1}^+ \to {3\over 2}_{1}^+ $     &0.16 &  0.95$\pm$.09  &26. &0.32 &2.4 & 1.9\\
 \hline
  ${5\over 2}_{1}^+ \to {1\over 2}_{1}^+ $    & 5.2 &  7.62$\pm$.24  & 20. & 4.3 & 7.5 & 8.6\\
 \hline
  ${5\over 2}_{1}^+ \to {3\over 2}_{1}^+ $  & 25. &  27.5$\pm$.9  & 1.1 & 27. & 27. & 20. \\
 \hline
  ${3\over 2}_{2}^+ \to {3\over 2}_{1}^+ $  & 19. &  26$\pm$24  & 14. & 15.7 & 17. & 15. \\
 \hline
  ${3\over 2}_{2}^+ \to {1\over 2}_{1}^+ $  & 11. &  20$\pm$-  & & & & \\
 \hline
  ${1\over 2}_{2}^+ \to {3\over 2}_{1}^+ $   & 23. &  10$\pm$6  & 32. & 31. & 2.4 & 6.5 \\
  \hline
  ${7\over 2}_{1}^+ \to {5\over 2}_{1}^+ $  & 0.7  &  1.6$\pm$1.3  &7.6 & 0.35 & 0.8 & 0.8 \\
  \hline
  ${7\over 2}_{1}^+ \to {3\over 2}_{1}^+ $  & 14. &  22.2$\pm$.19  & 11. & 22. & 25. & 23. \\
  \hline
  ${7\over 2}_{1}^+ \to {3\over 2}_{2}^+ $  & 0.16 &  1.52$\pm$.25  &     &    &    &  \\
  \hline
  ${7\over 2}_{2}^+ \to {3\over 2}_{1}^+ $  & 14.  &  1.6$\pm$.7  & 34. & 7. & 0.05 & 0.8 \\
\hline
  ${5\over 2}_{2}^+ \to {3\over 2}_{2}^+ $  & 0.7  &  4.  & 1.1 & 0.3 & 1.6 & 1.4 \\
\hline
  ${5\over 2}_{2}^+ \to {1\over 2}_{1}^+ $  & 12.&  25.7$\pm$.25  & 19. & 15. & 19. & 16. \\
\hline
  ${5\over 2}_{2}^+ \to {3\over 2}_{1}^+ $  &2.1 &  4.8$\pm$.5  & 3.8 & 11. & 5.1 & 7.3 \\
\hline
  ${9\over 2}_{1}^- \to {11\over 2}_{1}^- $  & 29. & 39$\pm$10  &&&&\\
\hline
  ${7\over 2}_{1}^- \to {11\over 2}_{1}^- $  & 2.7 &  $>$.49   &&&&\\
  \hline
\end{tabular}
\end{center}
\end{table}

 

For the odd Te neutron we use $g_l$=0.\,$\mu_n$ and $g_s$=-1.5\,$\mu_n$. 
The spin g-factor indicates some quenching from that of a free neutron. It should be noted that there is a wide range of $g_l$ and $g_s$ values that give a reasonable fit to the data. In considering the M1 operator one should keep in mind that for the special choice $g_b=g_l=g_s$ the operator \eqref{T-M1-IBFA} reduces to the operator for total angular momentum. Since this corresponds to a good quantum number the calculated B(M1) values vanish exactly for this choice. The relatively large B(M1) values thus requires a significant deviation from $g_b=g_l=g_s$ however the data do not allow an accurate determination of these parameters and a variation of the parameters with 30\% is possible without significantly spoiling the agreement.

Experimental B(E2) values for transitions between negative parity states are compared to the results of our calculations in Tables \ref{tbl:125Xe-BE2-wu}--\ref{tbl:131Xe-BE2-Wu}. In the tables we have expressed the values in terms of Weisskopf units where in this mass region for B(E2) values 0.37\,e$^2$b$^2$=100\,Wu while for B(M1) values the equivalence is given by 1.79\,$\mu_n^2$=0.0197\,e$^2$fm$^2$=1000\,mWu.

As can be seen from Tables \ref{tbl:123Te-BE2-Wu} -- \ref{tbl:129Xe-BE2-Wu} our results are in good agreement with the data for E2 transitions. The agreement with the data is of similar quality as that of other calculations where it should be noted that we have not optimized the parameters for each nucleus but rather used one set for all. The E2 transitions do not show a clear pattern that allows for the arrangement of the levels into bands. This is partly due to the fact that there are several single-particle levels important for the low-lying states and partly due to the fact that the even-even cores do not show strong collectivity.

Measured B(M1) values are only available for some transitions in $^{123}Te$ (\tabref{123Te-BE2-Wu}) and $^{125}Te$ (\tabref{125Te-BE2-Wu}). The calculation shows some large discrepancies for $^{123}Te$ while for $^{125}Te$ the over-all agreement is rather good, considerably better than for the calculation of ref.~\cite{J99}. The poorer agreement for M1 transitions could be due to the fact that for the M1-operator in IBFA higher-order terms are more important than for the E2-operator due to the fact that the M1 transitions are not collective.


\section{Spectroscopic factors}\seclab{SPT}

The operator for single particle transfer in the model is based on the microscopic interpretation of the structure of the bosons as fermion pairs. In one particle transfer the (generalized) seniority of the state may increase or decrease with one unit for pick-up (decreasing the number of target nucleons in for example a (p,d) reaction) as well as for stripping (such as (d,p) where the number of target nucleons in increased) reactions.
The general structure of the single particle transfer operator for stripping reactions where the odd fermion is a particle rather than a hole, can be decomposed as
\beq
T^\dagger_j= A^\dagger_j + B^\dagger_j\;, \eqlab{Td}
\eeq
where the first term is boson number conserving while the second term changes it by one unit. Since, by definition, the quasi-particle operator $a^\dagger_j$ increases the generalized seniority of the state by one unit and $s^\dagger d a^\dagger_j$ decreases it~\cite{Pittel}, the first term in \eqref{Td} is written as
\bea
A^\dagger_j &=& {1 \over K_j^-} \Big\{ u_j a^\dagger_j  \\
  &-& \sqrt{10\over 2j+1}  {\sqrt{N_\nu}\over N} v_{j} \sum_{j'} {\beta_{j',j}\over \sqrt{\cal N}_\beta} \Big[s^\dagger \tilde{d} a^\dagger_{j'} \Big]^j \Big\}  \nonumber \;.
\eea
Similarly the expression for the second, boson number changing, term in \eqref{Td} is
\bea
B^\dagger_j &=& {1 \over K_j^+} \Big\{ {v_j\over \sqrt{N}} s^\dagger \tilde{a}_j \\
  &+& \sqrt{10\over 2j+1}  {\sqrt{N_\nu}\over N} u_{j} \sum_{j'} {\beta_{j',j}\over \sqrt{\cal N}_\beta} \Big[d^\dagger \tilde{a}_{j'} \Big]^j \Big\} \nonumber \;.
\eea
The coefficients are defined as ${\cal N}_\beta= \sum_{j,j'} \beta_{j',j}^2$ where the normalization constants $K_j^\pm$ are chosen such that
\bea
\sum_j \langle odd (A+1) | A^\dagger_j| even (A) \rangle^2 &=& (2j+1) u^2_j \;,
\nonumber \\
\sum_j \langle even (A+2) | B^\dagger_j | odd (A+1) \rangle^2 &=& (2j+1) v^2_j \;.
\eea
Very similar formulas can be written for the case that the odd-fermion degree of freedom coupled to the bosons is hole-like in stead of particle like. For pick-up reactions the adjoint operators are used.


\begin{table}[htb!]
   \caption{ \tablab{123Te-Specf}
Spectroscopic factors for one neutron transfer from the ground state of $^{122}$Te and $^{124}$Te to various excited states in $^{123}$Te are compared with data, taken from ref.~\cite{V00} for $^{122}$Te(d,p)$^{123}$Te while from from ref.~\cite{Ohya94} for $^{124}$Te(d,t)$^{123}$Te and $^{124}Te$($^{3}He$,$\alpha$)$^{123}$Te.}.
\begin{tabular}
{|c||c|c|c||c|c|c|c||} \hline
&\multicolumn{3}{c||}{IBFM} &\multicolumn{4}{c||}{Data for levels in $^{123}$Te}
\\ \hline
j & Stripp. & E$_{\rm th}$ & Pickup & (d,p) & E$_{\rm exp}$ & (d,t) & ($^{3}$He,$\alpha$)
\\ \hline
$h_{11/2}$ & .41  &  .183& .58  &  .30  &  .247 & .48 & .28 \\\hline
$s_{1/2}$  & .45  &  .0  & .45  &  .39  &  .0   & .55 & .61\\\hline
$s_{1/2}$  & .05 &  .486& .05 &    .07 &  .599 &     &  \\\hline
$d_{3/2}$  & .33  &  .155& .38  &  .51  &  .159 & .43 & .43 \\\hline
$d_{3/2}$  & .00 &  .512& .00 &   .02 &  .440 &     &  \\\hline
$d_{3/2}$  & .11  &  .594& .12  &  .07 &  .688 & .16 &  \\\hline
$d_{5/2}$  & .13  &  .506& .51  &  .06 &  .490 & .20  & .48 \\\hline
$d_{5/2}$  & .04 &  .593& .16  &  .05 &  .688 & .10 &  .13 \\\hline
$d_{5/2}$  & .02 &  .720& .08 &  .07 &  .894 &  .33    &  \\\hline
$d_{5/2}$  & .00  &  .949& .00 &  .00   & 1.068 &     &  \\\hline
$g_{7/2}$  & .05 &  .496& .66  &  .02 &  .489 &     & .36 \\\hline
$g_{7/2}$  & .01 &  .787& .14  &  .01 &  .697 &     &  \\\hline
\end{tabular}
\end{table}

\begin{table}[htb!]
   \caption{ \tablab{125Te-Specf}
Spectroscopic factors for one neutron transfer from the ground state of $^{124}$Te and $^{126}$Te to various excited states in $^{125}$Te are compared with the data~\cite{J11}.}.
\begin{tabular}
{|c||c|c|c||c|c|c|c||} \hline
&\multicolumn{3}{c||}{IBFM} &\multicolumn{4}{c||}{Data for levels in $^{125}$Te}
\\ \hline
j & Stripp. & E$_{\rm th}$ & Pickup & (d,p) & E$_{\rm exp}$ & (d,t) & ($^{3}$He,$\alpha$)
\\ \hline
$h_{11/2}$ & .54  &  .106&   .45 &.31&  .145 &.51 & .42  \\\hline
$s_{1/2}$  & .45  &  .000&   .46 &.42&  .000 &.70 & .65  \\\hline
$d_{3/2}$  & .37  &  .038&   .42 &.46&  .035 &.62 & .60  \\\hline
$d_{3/2}$  & .01  &  .440&   .01 &   &  .440 &.03 &  .09   \\\hline
$d_{3/2}$  & .07  &  .570&   .08 &&  .521 &.02  & .01  \\\hline
$d_{5/2}$  & .00  &  .426&   .00 &   &  .464 &.01  &  -    \\\hline
$d_{5/2}$  & .02  &  .707&   .35 &.06&  .671 &.23  & .26  \\\hline
$d_{5/2}$  & .02  &  .817&   .42 &   & 1.053 &.17  & .20     \\\hline
$g_{7/2}$  & .05 &  .401&   .71 &   &   .444&.02 & .05         \\\hline
$g_{7/2}$  & .01 &  .691&   .12 &   &  .639 &.24   & .47 \\\hline
$g_{7/2}$  & .00 &  .920&   .00&.06 &  .641& .38  & -    \\\hline
\end{tabular}
\end{table}

\begin{table}[htb!]
   \caption{ \tablab{127Te-Specf}
Spectroscopic factors for one neutron transfer from the ground state of $^{126}$Te and $^{128}$Te to various excited states in $^{127}$Te are compared with the data~\cite{A127,H5}.}. 
\begin{tabular}
{|c||c|c|c||c|c|c||} \hline
&\multicolumn{3}{c||}{IBFM} &\multicolumn{3}{c||}{Data for levels in $^{127}$Te}
\\ \hline
j & Stripp. & E$_{\rm th}$ & Pickup & Pickup & E$_{\rm exp}$ & (d,p)
\\ \hline
$h_{11/2}$ & .41  &  .095&.57 &.45&  .088 &.23  \\\hline
$s_{1/2}$  & .46  &  .058&.47 &.75&  .061 &.20  \\\hline
$s_{1/2}$  & .02  &  .756&.02 &&  .622 &.00  \\\hline
$d_{3/2}$  & .43  &  .000&.49 &.62&  .000 &.24  \\\hline
$d_{3/2}$  & .00  &  .520&.00 &.01&  .501 &.01  \\\hline
$d_{3/2}$  & .02  &  .712&.02 &&  .762 &.00  \\\hline
$d_{5/2}$  & .00  &  .494&.00 &.00&  .473 &.00  \\\hline
$d_{5/2}$  & .03  &  .838&.67 & .28&  .782 &  \\\hline
$d_{5/2}$  & .00  &  .920&.18 & .14& 1.14 & .01  \\\hline
$g_{7/2}$  & .05  &  .751&.75 & .07&  .685 & .00  \\\hline
$g_{7/2}$  & .00  &  .859&.10 & .49&  .926 & .03  \\\hline
\end{tabular}
\end{table}

\begin{table}[htb!]
   \caption{ \tablab{129Te-Specf}
Spectroscopic factors for one neutron transfer from the ground state of $^{128}$Te and $^{130}$Te to various excited states in $^{129}Te$ are compared with the data~\cite{H07}.}.
\begin{tabular}
{|c||c|c|c||c|c|c||} \hline
&\multicolumn{3}{c||}{IBFM} &\multicolumn{3}{c||}{Data}
\\ \hline
j & Stripp. & E$_{\rm th}$ & Pickup & (d,p) & E$_{\rm exp}$ & (d,t)
\\ \hline
$h_{11/2}$ & .46  &  .091&.54 &.18&  .105 &.25  \\\hline
$s_{1/2}$  & .43  &  .189&.47 &.20&  .179 &.26 \\\hline
$s_{1/2}$  & .03  &  .814&.03 &.00&  .773 &-  \\\hline
$d_{3/2}$  & .38  &  .000&.57 &.33&  .000 &.27 \\\hline
$d_{5/2}$  & .00  &  .539&.12 &.00&  .544 &.00  \\\hline
$d_{5/2}$  & .02  &  .565&.76 & .01&  .633 &-  \\\hline
$g_{7/2}$  & .06  &  .765&.81 & .01&  .812 &.01  \\\hline
$g_{7/2}$  & .00  &  .901&.06 & .00&  .865 &.00  \\\hline
\end{tabular}
\end{table}

\begin{table}[htb!]
   \caption{ \tablab{131Te-Specf}
Spectroscopic factors for one neutron transfer from the ground state of $^{130}$Te and $^{132}$Te to various excited states in $^{131}$Te are compared with the data~\cite{J84}.}.
\begin{tabular}
{|c||c|c|c||c|c||} \hline
&\multicolumn{3}{c||}{IBFM} &\multicolumn{2}{c||}{Data}
\\ \hline
j & Stripp. & E$_{\rm th}$ &  Pickup & (d,p) & E$_{\rm exp}$
\\ \hline
$h_{11/2}$ & .33  &  .173&.65&.15 & .182    \\\hline
$s_{1/2}$  & .45  &  .286&.50&.13 & .296   \\\hline
$s_{1/2}$  & .00  &  .997&.00&.00 & 1.041   \\\hline
$d_{3/2}$  & .39  &  .000&.59&.5 & .000   \\\hline
$d_{5/2}$  & .02  &  .624&.89&.00 &  .642   \\\hline
$g_{7/2}$  & .03  &  .920&.48&.00 & .943   \\\hline
\end{tabular}
\end{table}

\begin{table*}[htb!]
   \caption{ \tablab{Xe-Specf}
Calculated spectroscopic factors for one neutron transfer from the ground state of $^{124}$Xe, $^{126}$Xe, $^{128}Xe$ and $^{130}Xe$ to various excited states in $^{125}$Xe , $^{127}$Xe, $^{129}$Xe and $^{131}$Xe .}
\begin{tabular}
{|c||c|c|c||c|c|c||c|c|c||c|c|c||} \hline
&\multicolumn{3}{c||}{ $^{125}$Xe}& \multicolumn{3}{c||}{ $^{127}$Xe}& \multicolumn{3}{c||}{ $^{129}$Xe} &\multicolumn{3}{c||}{ $^{131}$Xe}
\\ \hline
j & Stripp. & E$_{\rm th}$ &  Pickup & Stripp. & E$_{\rm th}$ &  Pickup & Stripp. & E$_{\rm th}$ &  Pickup & Stripp. & E$_{\rm th}$ &  Pickup
\\ \hline
$h_{11/2}$   & .25 &  .303&.64 &    .40 & .309&.57&       .61& .213&.61&      .32& .127&.63   \\\hline
$s_{1/2}$    & .36 &  .000&.38 &    .40 & .000&.42&       .42& .000&.41&      .43& .089&.43   \\\hline
$s_{1/2}$    & .10 &  .412&.10 &    .03 & .613&.04&       .03& .577&.04&      .04& .630&.04  \\\hline
$s_{1/2}$    & .00 &  .835&.00 &    .04 & .799&.03&       .02& .745&.03&      .02& .798&.02   \\\hline
$d_{3/2}$    & .23 &  .094&.29 &    .19 & .133&.23&       .37& .029&.41&      .42& .000&.46   \\\hline
$d_{3/2}$    & .00 &  .390&.00 &    .06 & .371&.07&       .00& .317&.00&      .00& .381&.00   \\\hline
$d_{3/2}$    & .16 &  .559&.18 &    .05 & .591&.07&       .05& .544&.07&      .04& .613&.03   \\\hline
$d_{5/2}$    & .10 &  .385&.46 &    .00 & .362&.00&       .00& .315&.00&      .00& .348&.00   \\\hline
$d_{5/2}$    & .02 &  .488&.10 &    .00 & .553&.18&       .00& .638&.04&      .00& .773&.02   \\\hline
$d_{5/2}$    & .03 &  .614&.17 &    .00 & .702&.12&       .01& .792&.35&      .03& .934&.55   \\\hline
$d_{5/2}$    & .00 &  .774&.00 &    .01 & .826&.02&       .01& .905&.28&      .01&1.026&.17   \\\hline
$g_{7/2}$    & .03 &  .324&.55 &    .02 & .350&.41&       .01& .611&.10&      .00& .706&.02   \\\hline
$g_{7/2}$    & .01 &  .650&.11 &    .01 & .647&.30&       .03& .754&.47&      .04& .853&.63   \\\hline
$g_{7/2}$    & .00 &  .675&.07 &    .00 & .718&.00&       .00& .865&.02&      .00& .950&.04   \\\hline
\end{tabular}
\end{table*}


The calculated values for spectroscopic factors are compared with data for $^{123,125,127,129,131}$Te~\cite{V00,J11,H5,H07,J84} in \tabref{123Te-Specf} till \tabref{131Te-Specf}. For the calculations the program SPEC~\cite{Ho0} is used. The obtained results for the neighboring $^{125-131}$Xe isotopes, for which there are -unfortunately- no experimental data available, are given in \tabref{Xe-Specf}.

The spectroscopic factors depend sensitively on the occupation probabilities of the single particle levels ($v_j^2$) in as given in \tabref{BCS-param-Te} and \tabref{BCS-param-Xe} which also enter in the calculation of excitation energies and electromagnetic transitions. A good example of the nice agreement obtained can be seen in the results for $^{123}$Te (\tabref{123Te-Specf}). In the calculation for the $11/2^-$ levels the spectroscopic factor for Stripping is smaller than for Pickup due to the fact that $v_{11/2}^2>0.5$, in nice agreement with the data. 
Also for the other single particle levels the trends are reproduced correctly. For the $3/2^+$ levels the calculation predict almost vanishing values for the second level, again in agreement with the observations.

In $^{127}$Te some discrepancies are observed for the $5/2^+$ and $7/2^+$ levels where the spectroscopic factor for Pickup is predicted to be large while the data indicate much larger values for Stripping. This probably indicates that the occupancies for the $d_{5/2}$ and the $g_{7/2}$ levels has been taken too large.

\section{Conclusions}

In the present work systematic calculations in the IBFA model for of odd-mass $^{125-131}$Xe and $^{123-131}$Te are described by coupling the degrees of freedom of a single fermion to the even-even core as described in~\cite{Pascu10}. We obtained a rather detailed agreement for the excitation energies of all isotopes in consideration using a fixed interaction strength for the whole region where only the energies and occupation probabilities of the single particle levels coupled to the system of bosons is are varying across the mass range. The same occupation probabilities also enter in the operators for electromagnetic transitions and more explicitly in the expressions for single-particle-transfer probabilities. The latter form therefore a sensitive test of the same parameters that enter in the calculation of excitation energies.

To conclude, our results show that an accurate description can be obtained for an extended region of isotopes in the IBFA model using interaction strengths that are fixed over the whole mass region which single particle parameters that show a systematic variation.

\begin{acknowledgments}
The present work has been performed with financial support from the University of Groningen (RuG) and the Helmholtzzentrum fuer Schwerionenforschung GmbH (GSI), Darmstadt. We wish to thank D. Bucurescu for discussions regarding this work.
\end{acknowledgments}


\end{document}